\documentstyle[editedvolume,psfig]{crckapb10} 

\begin{opening}
\title{Cosmology with redshift surveys of radio sources}
\author{Steve Rawlings}
\author{Katherine M. Blundell}
\author{Mark Lacy}
\author{Chris J. Willott}
\institute{Astrophysics,Department of Physics,Keble Road,Oxford OX1 3RH}
\author{Stephen A. Eales}
\institute{Department of Physics and Astronomy, University of Wales at
Cardiff, P.O. Box 913, Cardiff CF2 3YB}

\end{opening}

\runningtitle{Cosmology with redshift surveys of radio sources}

\begin{document}

\begin{abstract}{\small

We use the $K-z$ relation for radio galaxies to illustrate why 
it has proved difficult to obtain definitive cosmological results from 
studies based entirely on catalogues of the brightest radio sources, e.g.
3C. To improve on this situation we have been
undertaking redshift surveys of complete samples drawn from the 
fainter 6C and 7C radio catalogues. We describe these surveys, and
illustrate the new studies they are allowing. 
We also discuss our `filtered' 6C 
redshift surveys: these have led to the discovery of a 
radio galaxy at $z=4.4$, and are sensitive
to similar objects at higher redshift provided the space density of these
objects, $\rho$, is
not declining too rapidly with $z$. There is currently no direct evidence
for a sharp decline in the $\rho$ of radio galaxies for $z > 4$, a result
only barely consistent with the observed decline of flat-spectrum
radio quasars.
}
\end{abstract} 

% The \begin{document} command comes after the \end{opening}
% command.

\section{Introductory remarks}

The few column inches devoted to radio galaxies by Peebles (1993) reflects
a common view that they are objects  
of rather peripheral interest to cosmology.
Nevertheless, because of the ease with which
they can be found at high redshift, they have become popular objects to
study. Many ($\sim 100$) are now known at $z > 2$,
and the most distant of these (at $z=4.4$, Rawlings et al. 1996) 
is not far from the $z=4.9$ 
record for quasars. The days when radio galaxies were the only 
known galaxies at high redshift are, however, now over 
(Steidel et al. 1996).

Radio galaxies may still allow some 
special insights into questions of cosmological interest.
The most likely low-$z$ counterparts of the 
$z > 3$ population discovered by Steidel et al. are the spheroids
of early-type spirals (Trager et al. 1997). Radio galaxies, 
on the other hand, at low- and intermediate-$z$
seem to be associated exclusively with giant 
elliptical galaxies. By studying them at high $z$, one can hope to  
learn about the formation and evolution of massive ellipticals, and perhaps 
constrain cosmological parameters 
(e.g. the study of 53W091 by
Dunlop et al. 1996). Although other techniques for finding high-$z$
ellipticals are beginning to show promise (e.g. Graham \& Dey
1997), radio galaxies may continue
to trace the most massive galaxies and clusters at 
high redshifts.

\section{The need for redshift surveys fainter than 3C}

The near-IR ($K-z$) Hubble Diagram for radio galaxies 
{\it is} discussed by Peebles (1993), and acts as a good focus for an
examination of the possible pitfalls of cosmological 
studies based entirely on surveys of the brightest radio sources. 
The $K-z$ relation for 3C galaxies is included in Fig. 1.

\begin{figure}
\vspace{-1cm}
\centerline{
\psfig{figure=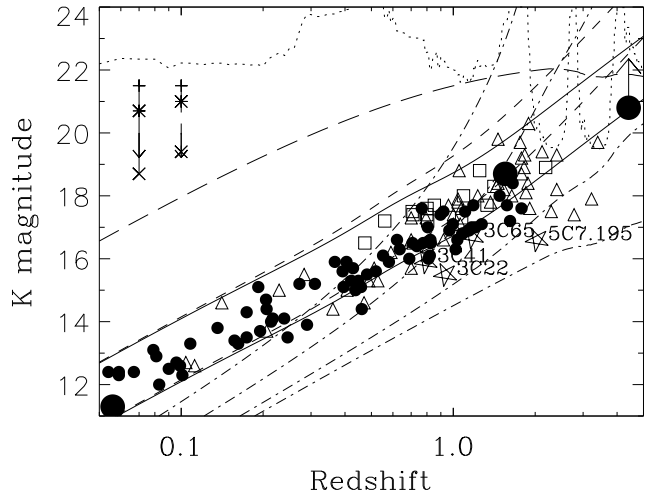,width=8.5cm}
}
\vspace{-0.5cm}

\caption[ki]{ 
The $K-z$ relation for radio galaxies from 3C (filled circles),
6C (triangles), 7C (squares), confirmed `red quasars'
(stars), and as large filled circles: Cyg~A, $z=0.056$; 53W091, $z = 1.55$;
6C0140+326, $z=4.41$. The solid lines bracket non-evolving
({\it k}-corrected) ellipticals with luminosities between $M_{*}$
and $M{*}-2$ (a typical BCG value) for $\Omega_{\rm M}=1$; the
dashed lines are for $\Omega_{\rm M}=0$. The vectors show the expected
brightening with look-back time due to
passive stellar evolution from $z=0$ (`+') to
$z=1$ (`*') to $z=4.4$ (`$\times$'): $\Omega_{\rm M}=1$, left;
$\Omega_{\rm M}=0$, right. The dot-dash lines show the loci of 3C
quasars reddened by $A_{V}=0,2,7,15$ (bottom to top). The long-dash line
shows the loci of the scattered light from a 3C quasar assuming 
that the scattering is by optically thick dust with a covering factor
$\approx 0.01$, and a $\lambda^{-2}$ dependence for the 
scattering efficiency (e.g. Ogle et al.
1997). The dotted line shows the expected contamination due to 
narrow emission lines (for 3C objects), 
with line ratios taken from composite spectra
of high-$z$ radio galaxies and, at low redshift, from 
an unpublished spectrum of Cyg~A.

} 
\end{figure}

We consider first the implications of assuming that 
the $K$ magnitudes of all the 3C objects plotted in Fig. 1 
are dominated by starlight  
(quasars, i.e. objects with broad optical emission lines, have been
excluded). As discussed by Lilly \& Longair (1984),
the highest redshift 3C galaxies are about 1 magnitude
more luminous than their low $z$ counterparts. This
can be interpreted in one of two ways. Either the 3C galaxies are, at all
redshifts, similar mass galaxies which brighten
systematically with look-back time because of passive evolution of their
stellar populations. Or, at $z  
\stackrel{>}{_\sim} 1$, the 3C galaxies are
exclusively ultra-massive brightest cluster galaxies (BCGs) 
like the host of Cyg~A (Fig. 1), whereas
at low redshifts they are typically much less massive elliptical galaxies.
Some authors argue that both effects are operating
(Best, Longair \& R\"{o}ttgering 1997).

But, are the $K$ magnitudes of 3C radio galaxies
dominated by old stars? The predicted $K$ values of various
non-stellar sources are shown in Fig. 1, these are:

\begin{itemize}

\item {\bf Emission lines.} Their huge influence on the 
$K-z$ relation for $z > 2$ is well documented (Eales \& Rawlings 1993,1996), 
but they also have important effects at lower $z$. In Fig. 2 we present
a $K-$band spectrum of the $z=1.1$ radio galaxy 
3C368. Diffuse red continuum is seen in this spectrum 
(particularly beyond $2.4 ~ \mu{\rm m}$ where CO absorption dims the 
foreground M star), but strong [SIII] and Helium lines contribute 
$\approx25 \%$ of the spatially-extended light, having a strong influence on 
the appearance of the  `galaxy' (Stockton, Ridgway \& Kellogg 1996).
Fig. 1 shows that, at $z < 2$,
this contaminant can account for at most $\sim 10 \%$ of the 
total $K-$band light of the average 3C galaxy.

\begin{figure}
\centerline{
\psfig{figure=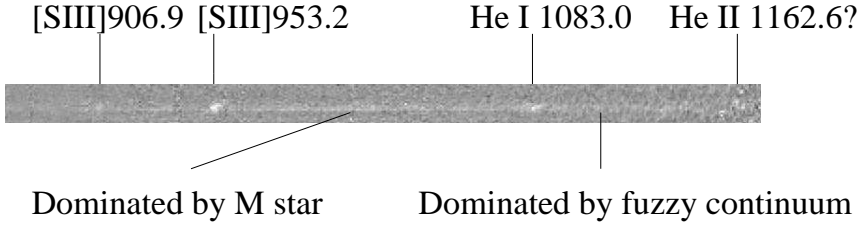,angle=-90,width=11.5cm}
}
\caption[ki]{\label{fig} 
A $K-$band spectrum of 3C368 ($z = 1.132$)
obtained at the UKIRT; the spatial dimension is 20 pixels ($\approx 24 ~ \rm
arcsec$) along PA=$8^{\circ}$ with North roughly upwards. 
The [SIII] lines are resolved both spatially (size $\approx 3 ~ \rm
arcsec$), and in velocity ($\approx 500 ~ \rm km ~ s^{-1}$ shift
N-S).

} 
\end{figure}

\item {\bf Buried quasar nuclei.} The broad emission lines discovered
in the near-IR spectra of 3C22 (e.g. Rawlings et al. 1995)
and 5C7.195 (Willott et al., these proceedings) 
demonstrate that lightly-reddened ($A_{V} \sim 1$) quasars can
masquerade as narrow-line radio galaxies if classification is based
solely on optical spectroscopy. 
Optical {\sc HST} images of high-$z$ 3C radio galaxies (Best et al. 1997)
probe to $A_{V} \sim 7$, compared with the $A_{V} \sim 2$ possible with near-IR
spectroscopy, but still reveal a low fraction of
lightly reddened quasars (e.g. 3C22). 
Fig. 1 shows that buried nuclei
can still contribute significantly to the $K-$band light
provided $A_{V} \sim 7$. By making the 
first 3.5-$\mu \rm m$ detections of four $z \sim 1$ 3C radio galaxies,
Simpson, Rawlings \& Lacy (1997) have recently shown that buried nuclei
can be an important contributer at $K$, providing $\sim10\%$ of the light
in one case where $A_{V} > 7$, and $\approx50-100\%$ in the cases
of 3C22, 3C41 and 3C65.

\item {\bf Scattered quasar light}. Fig. 1 shows that dust scattering of light
from a buried quasar, as is now established in Cyg~A (Ogle et al. 1997),
is also likely to contribute at most 
only $\sim 10 \%$ of the total $K-$band light
of $z < 2$ radio galaxies. Such a low level of contamination
has been confirmed by the detection of small
(at most a few per cent) $K-$band polarizations for $z \sim 1$ 3C radio 
galaxies (Leyshon \& Eales 1997). 

\end{itemize}

To this list of non-stellar contaminants one can also add 
a further, and rather puzzling, source of red radio-aligned light 
(e.g. Dunlop \& Peacock 1993) but again this rarely seems to contribute 
at more than the $\approx 10 \%$ level, and then only in the
most extreme (3C) radio sources
(Eales et al. 1997).

To conclude, with certain exceptions (e.g. 3C22, 3C41, 3C65),
the resolved light profiles of $z \sim 1$ radio galaxies
(e.g. Best et al. 1997), and the low 
($\stackrel{<}{_\sim} 10 \%$) levels of potentially spatially resolved 
contaminants means that the stellar luminosities of
3C radio galaxies {\it are} probably higher at high redshifts.
However, this may still be a selection effect associated with extreme
radio luminosity: the brightest radio sources probably require 
both the most powerful jets {\it and} the densest gaseous environments
(e.g. Rawlings \& Saunders 1991) -- environments which are only 
associated with the most massive brightest cluster galaxies (BCGs).

\section{The 6C and 7C redshift surveys}

We decided several years ago to seek 
redshifts for low radio-frequency complete samples significantly
fainter than 3C. Details of these 6C and 7C samples are given
in Table 1, and the improved
coverage of the 151-MHz luminosity ($L_{151}$), $z$ plane they provide is
discussed by Blundell et al. (these
proceedings). The 6C sample has virtually complete redshift
information. About 25\% of the 7C radio sources are associated 
with quasars (see Willott et al., these proceedings), and a further
65\% have narrow emission lines. These features of the
7C sample again ensure a high redshift completeness 
($\approx 90 \%$ in the 7C-1 region where spectroscopy
has been completed) but are of course a 
mixed blessing since the emission lines are almost certainly an
indication of the AGN activity which, as noted in \S2, compromises
studies of the 3C sample. In 7C, however, the AGN contamination 
is at a lower level: the radio/optical correlation for radio quasars 
(Serjeant et al. 1997),
and associated narrow-line/radio correlations for radio sources
(e.g. Rawlings \& Saunders 1991), suggest that the loci of the
three non-stellar contaminants in Fig. 1 should scale as $S_{151}^{p}$ with
$p \approx 0.6$, and these should have little influence on the $K-z$ relation,
at least for $z < 2$.

About $\approx 10\%$ of the 7C sources will lack
secure redshifts even after completion of our spectroscopic 
campaign. These objects have already been imaged in several 
near-IR colours, and we find that most have the spectral 
energy distributions typical of
galaxies with $z \sim 1.5$, evolved stellar populations, 
and weak/absent emission lines.
In other words, these galaxies are very similar to 
53W091 (Dunlop et al. 1996), and, with large-telescope follow-up,
may together provide an even stronger constraint on the age of the 
high-$z$ Universe. 

\begin{table}

\caption{Details of the redshift surveys}

\begin{tabular}{lrrrrr}

Survey     & Radio limit       & Sky area    &  No     & Redshifts   & Filter \\
           &                   & (sr)       &         &             &           \\
           &                   &            &         &             &  \\

3C         & $S_{151} > 12$    & 4.2    & 173     & 100\%       & None        \\
6C         & $4 > S_{151} > 2$ & 0.1    &  63     &  97\%       & None        \\
7C-1 (5C6) & $S_{151} > 0.5$   & 0.0065 &  39     &  90\%       & None        \\
7C-2 (5C7) & $S_{151} > 0.5$   & 0.0065 &  40     &  65\%       & \ddag None   \\
7C-3 (NEC) & $S_{151} > 0.5$   & 0.0086 &  54     &  70\%  & \ddag None        \\
           &                   &        &         &             &           \\
6C$^{*}$     & $2 > S_{151} > 1$ & 0.133  &  34     &  60\% & 
$\phantom{aaaa}$
\ddag $\alpha>1$,
$\theta < 15^{\prime \prime}$      \\
6C$^{**}$    & $S_{151} > 0.5$   & $\approx0.2$ & $\approx 100$ &  15\%  &
\ddag $\alpha>1$, $\theta < 10^{\prime \prime}$ \\

\end{tabular} 

 \parbox{\linewidth}{
\ddag This means that spectroscopy has yet to be attempted on all
members of these samples.

}

\end{table}

\section{Cosmology with the 6C and 7C surveys} 

Here, we will only discuss some of the implications from the 
$K-z$ relation from these new redshift surveys (other results are
presented by Blundell et al. and Willott et al., these proceedings).
$K-$band photometry of the 7C-1 and 7C-2 
galaxies has only just been completed, so we will focus 
our discussion on the (mainly 6C) data available at the time of the
Tenerife meeting (Fig. 1; see also Eales \& Rawlings
1996; Eales et al. 1997). 
Preliminary analysis of the new 7C data indicates that we will eventually 
have a much sounder statistical basis for our conclusions.

At $z \sim 1$ the separation of 3C and 6C/7C points 
in Fig. 1  indicates that
3C radio galaxies {\it are} brighter by virtue
of their extreme radio luminosity. It remains unclear whether this effect
is due to contaminant light sources or to the
3C objects being associated with more massive galaxies. 
However, considering the total spread in stellar luminosities
of radio galaxies there now seems little evidence for any evolution 
between $z \sim 0$ and $z \sim 1$. At either epoch the  
magnitudes of the radio galaxies are sandwiched between those
of an unevolved $M_{*}$ galaxy and an unevolved BCG. 
We are therefore led to two possibilities: either (i)
$\Omega_{\rm M}=1$ and the brightening with look-back time
due to passive stellar evolution is cancelled by the effects of 
accretion by mergers; or (ii) $\Omega_{\rm M} < 1$.
We plan to use high-resolution near-IR imaging to distinguish between these
possibilities.  

The curves of Fig. 1 indicate that one has to worry much more
about all the contaminant light sources in any interpretation
of the $K-z$ diagram for $z > 2$ .
It is not yet clear whether the large increase in dispersion
is indicative of a wide range of ages in radio galaxies for $z > 2$, and
thus whether they are being seen at times when they
were young and/or forming (Eales \& Rawlings 1996).

\section{Filtered 6C redshift surveys}

Our new 6C and 7C redshift surveys have gone a 
long way towards breaking the degeneracy between $L_{151}$ and $z$ 
in the study of radio galaxies, 
and have extended the redshift coverage by complete samples
to encompass the range $2 < z < 3$. They include, however, just 
2 objects at $z > 3$. This is largely a consequence 
of the limited sky area of these surveys. In tandem with our
complete sample work we have also been undertaking redshift surveys
of larger areas at flux levels comparable to that of the 7C survey.
To ensure optical follow-up
is confined to a manageable number of sources requires that we
filter out some large fraction ($>90\%$) 
of the sources using radio
selection criteria. The filtering criteria we have used for 
our 6C$^{*}$ and 6C$^{**}$ redshift surveys are given in Table 2.

Most $z > 3$ radio galaxies have been found from samples which exclude 
all the sources with radio spectral indices $\alpha$ flatter than
some critical value, and with radio angular sizes $\theta$ greater than
some critical value. Choosing these values involves a trade-off 
between the `efficiency' (i.e. the fraction of the sample which lies at
$z >  z_{\rm target}$), and the incompleteness (i.e. the fraction of the
$z > z_{\rm target}$ population which has been rejected by the 
filtering criteria). 
When $z_{\rm target}=2$ we can assess these factors by comparing the 6C$^{*}$
sample with a sub-set of the 7C survey which is matched in
$S_{151}$: this comparison suggests that about 80\% of the $z > 2$ 
population (meeting the $S_{151}$ criteria)
has been missed by 6C$^{*}$, mostly  
because of the $\alpha$ criterion, but that 6C$^{*}$ is twice as
efficient ($\approx40 \%$ {\it versus} $\approx20 \%$ for 7C) at finding
$z > 2$ radio galaxies.

When $z_{\rm target} \geq 3$ 
the lack of objects in the complete samples precludes a 
similarly direct assessment. However, since the $z>3$ radio sources will all
lie at the top of the radio luminosity function (RLF)
we can compare their $\alpha$ and $\theta$ properties
with those of the most radio-luminous 3C galaxies
(Fig. 3). Most objects have concave radio spectra like Cyg~A, so the 
{\it k}-correction means that an $\alpha > 1$ criterion should 
exclude only a minority of the $z > 3$ objects. 
A $\theta < 10 ~ \rm arcsec$ criterion requires a 
strong negative evolution of linear size with $z$ if most $z > 3$ objects
are to be retained. Statistically speaking there is good 
evidence for just such an 
evolutionary trend (e.g. Neeser et al. 1995; Blundell et al., these 
proceedings), but an intrinsic spread in $\theta$ means that 
at least some $\theta > 10 ~ \rm arcsec$ sources are already 
known at $z > 3$ (Fig. 3). 
We suspect, therefore, that the $\theta$ selection criteria
of our 6C$^{*}$ and 6C$^{**}$ surveys are likely to be just as severe 
causes of incompleteness as those due to the $\alpha$ criteria,
especially if $\Omega_{\rm M}=1$.
The influences of filtering criteria need to be 
considered very carefully, especially
if one is to use filtered samples in any analysis of 
the space density of high-$z$ radio galaxies.

\begin{figure}
\vspace{-1cm}
\centerline{
\psfig{figure=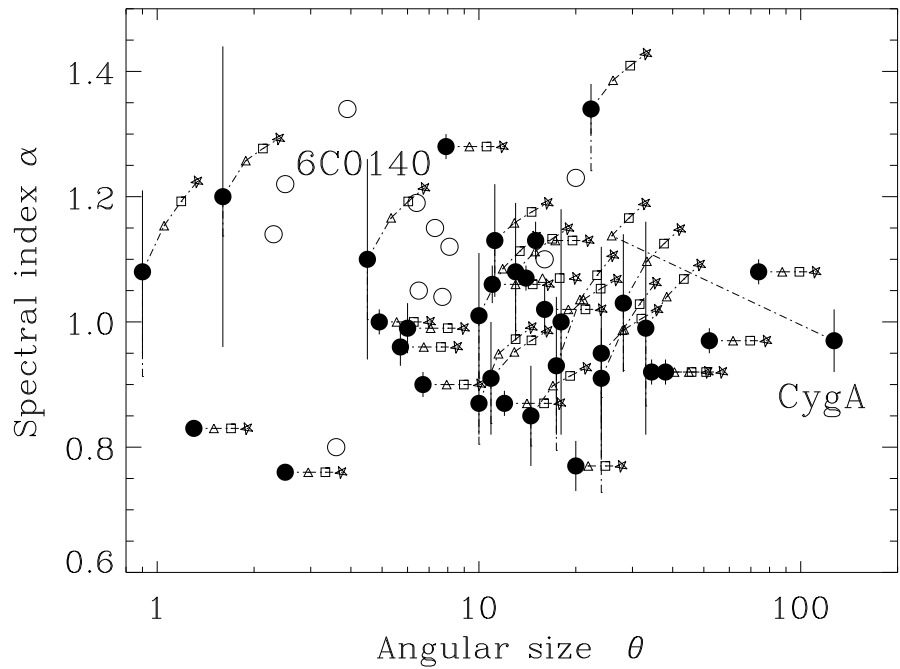,angle=90,width=12.5cm}
}
\caption[ki]{\label{fig} 
The radio spectral index $\alpha$ (evaluated at 
1 GHz), angular size $\theta$ (in arcsec) plane for
3C radio galaxies (filled circles, Cyg~A marked) 
and the $z > 3$ radio galaxies known prior to the 
Tenerife conference (open circles, 6C0140+326 marked).
For the 3C sources we have used their integrated 
radio spectra and observed angular sizes to predict the loci of similar radio
sources at $z = 3$ (triangles), $z = 4$ (squares) and $z = 5$ (stars);
we assume $\Omega_{\rm M}=1$.

} 
\end{figure}

Despite our concerns about incompleteness, the 6C$^{*}$ survey did lead to the
discovery of 6C0140+326 at a redshift of $4.41$ which is the most 
distant radio galaxy currently known (Rawlings et al. 1996).
Allowing for passive stellar evolution
the galaxy seems too faint to be either a well-formed giant elliptical or 
an unobscured star-forming elliptical in an $\Omega_{\rm M}=1$
cosmology (it has only a $K$ limit in Fig. 1). Either then
we invoke some complicated interplay between the dynamical status of the
galaxy, the age of its stellar population and/or the presence of dust, 
or, as argued in \S4, we consider models in which $\Omega_{\rm M}<1$.

\section{The space densities of high-redshift radio galaxies}

Another cosmological use for radio sources is in the
pinpointing of the rapid decline, or `redshift cut-off',
in the co-moving space density $\rho$ of massive 
(and hence radio-luminous) galaxies expected at $z \sim 5$
(e.g. Efstathiou \& Rees 1988).
By utilising all the data available at the time
(e.g. redshift surveys and source counts), 
Dunlop \& Peacock (1990) found some
evidence that steep-spectrum radio galaxies decline in $\rho$ beyond a 
peak at $z \sim 2.5$. However, given the many difficulties involved
with this work (e.g. radio {\it k}-corrections, redshift estimates, 
small number statistics) even the most hardened advocates of a global
redshift cutoff might concede that for the most radio-luminous
steep-spectrum population at least, the evidence is not yet 
conclusive. The evidence for a significant decline in 
$\rho$ for flat-spectrum radio quasars 
appears to be considerably firmer (Dunlop \& Peacock 1990), indeed
possibly now incontrovertible (Shaver et al. 1996).

\begin{figure}
\vspace{-1cm}
\centerline{
\psfig{figure=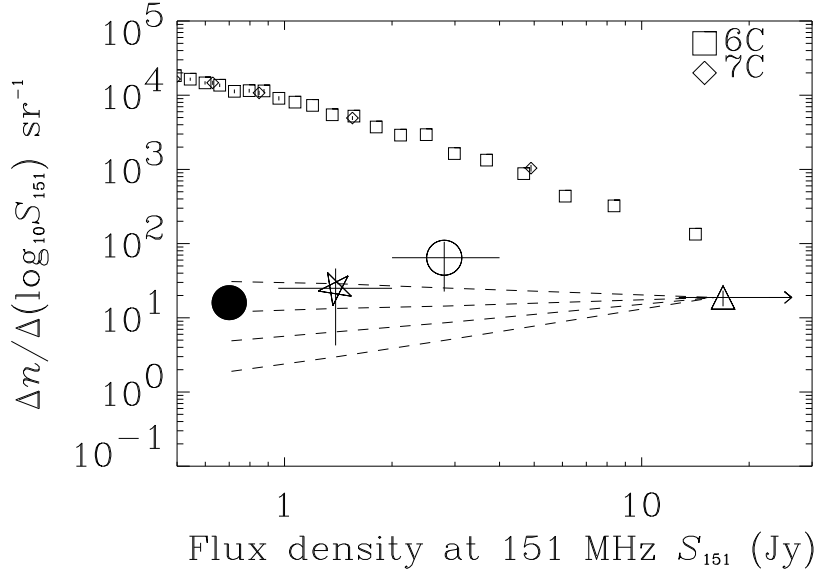,width=12.5cm}
}
\vspace{-0.5cm}
\caption[ki]{\label{fig} 
Areal density of high-$z$ radio sources
(in the top decade of the RLF)
as a function of $S_{151}$. 
The upper data show the total 6C and 7C source counts. The lower
points represent: 3C radio sources at $z > 1.3$ (triangle);
6C sources at $z > 3$ (open circle); 6C$^{*}$ sources at
$z > 4$ (star); and a point corresponding to 1 radio
source at $z > 5$ in 6C$^{**}$ (filled circle). The dashed lines
illustrate how this depends on the strength of the
redshift cut-off from {\it (upper)} no cut-off to {\it (lower)} a
cut-off as strong as that seen in the flat-spectrum population 
(Shaver et al. 
1996).
} 
\end{figure}

We have begun our analysis by asking a very simple question.
If we 
concentrate on the most radio-luminous galaxies,
do we see any direct evidence for a sharp decline in $\rho$? Fig. 4
shows that as yet we do not: our discovery of
6C0140+326 (and the fact that the other known $z > 4$
radio galaxy, Lacy et al. (1994), was found from a 0.2 sr survey), 
implies that $\rho$ is roughly constant over the redshift
range $1.3 \leq z \leq 4.5$. 
These results are only barely consistent with the rapid decline in
$\rho$ for flat-spectrum radio quasars (Fig. 4): a full
understanding of the suspected gravitational lensing of both the known $z > 4$ 
radio galaxies (Lacy et al. 1994, Rawlings et al. 1996)
may help bring these results into closer accord.
Fig. 4 gives us some hope that the 6C$^{**}$ survey will 
include at least one radio galaxy at $z > 5$. 

%\begin{quote}
%{\small
%\begin{verbatim}

%\end{verbatim}}
%\end{quote}

\end{document}